# Perfect valley filter in strained graphene with single barrier region

Can Yesilyurt,[1] Seng Ghee Tan,[1,2] Gengchiau Liang,[1] and Mansoor B. A. Jalil [1,a]

[1]*Electrical and Computer Engineering, National University of Singapore, Singapore 117576, Republic of Singapore*
[2]*Data Storage Institute, Agency of Science, Technology and Research (A\* Star), Singapore 138634, Republic of Singapore*

We present a single barrier system to generate pure valley-polarized current in monolayer graphene. A uniaxial strain is applied within the barrier region, which is delineated by localized magnetic field created by ferromagnetic stripes at the region's boundaries. We show that under the condition of matching magnetic field strength, strain potential, and Fermi energy, the transmitted current is composed of only one valley contribution. The desired valley current can transmit with zero reflection while the electrons from the other valley are totally reflected. Thus, the system generates pure valley-polarized current with maximum conductance. The chosen parameters of uniaxial strain and magnetic field are in the range of experimental feasibility, which suggests that the proposed scheme can be realized with current technology.

## I. Introduction

The discovery of graphene[1] has attracted much attention due to its unique properties such as zero-gap band structure and linear energy-momentum dispersion relation near the Dirac points, situated at the corners of the Brillouin zone of the graphene's reciprocal lattice structure[1-3]. The two Dirac points exhibit two different inequivalent characteristics, which are labelled as the $K$ and $K'$ valleys[2]. This spin-like valley degree of freedom of graphene has opened up a new avenue in nanoelectronic applications, which is the so-called valleytronics[4]. As a matter of course, the ability to generate valley-polarized current becomes an essential requirement in realizing the valleytronic applications. One way to break the valley symmetry in graphene is by applying a strain to it. The flexible structure of graphene makes it amenable to the application of strain[5,6]. It has been shown that the strain field breaks the symmetry of the coupling strengths of the honeycomb lattice and hence induces an effective valley-dependent potential[6-8]. The breaking of the valley degeneracy enables us to utilize this effect as a means to attain a valley filter and polarizing function. In the literature, several valley filter applications have been proposed[9-13]. However, they have some limitations such as practical difficulties and low filter efficiency, as reported in Ref. 11 and Ref. 14, respectively. One of the prerequisites for the generation of the valley polarized current is the breaking of inversion or time-reversal symmetry. This has been achieved by using the magnetic field in most of the proposed valley filters. Additionally, the generation of valley polarized current may also be achieved by breaking the time reversal symmetry with time-dependent fields instead of a magnetic field in non-equilibrium graphene systems, as in Ref. 15. Another alternative is to utilize line defects on graphene to generate valley polarized current[16]. A proposed valley filter which is similar to our set-up comprises of two different region, i.e., a region of strained graphene substrate which acts as a valley separator, followed by an extractor region with a local perpendicular magnetic field[9,17]. In general, the valley polarization is not close to 100%[17], but by matching the strain and magnetic fields, ideal valley polarization can be approached[9]. In the two-barrier structure, incident electrons are polarized in angular space at the first barrier interface, before being filtered out at the second barrier interface. This dual process provides quite a robust valley filtering process by means of strain application. However, our present scheme has two distinct advantages: (i) the valley filters in Ref. 9 and 17 have the disadvantage of having low conductance when it reaches high valley polarization; (ii) our present scheme also provides a simpler configuration consisting of a single barrier region that integrates both the valley separation and extraction steps. Another proposal[18] utilizes the different pseudospin structure of the valleys in bilayer graphene to generate valley-polarized currrent, but again, the conductance shows a decreasing trend when the valley polarization is approaching its maximum. Thus, the present valley filter proposals face a general problem of maintaining a high conductance while at the same time achieving a high valley-polarization of current, the twin requisites for an efficient valley filter.

---

[a] elembaj@nus.edu.sg

**II. The valley filter model**

In this work, we present a highly-efficient (> 99%) valley filter based on strain engineered monolayer graphene, with a single barrier region. The use of potential barriers is one of the common ways to generate a filter function. However, any kind of barrier affects the energy under the barrier region, and hence, the momentum of electrons in the propagation direction. Based on the wave nature of the electron transport, this induces a refraction at the barrier interface, and reflection for the electrons whose incident angle is larger than the critical angle. Consequently, there is a decrease in the conductance of system due to the reflected electrons. The present filter scheme aims to overcome this issue and realize a filter operation with maximum conductance, and zero reflection of electrons with the desired valley degree of freedom. Furthermore, we seek to achieve this ideal behavior using parameters for strain and the magnetic field which are practically accessible. Basically, the filtering method is based on using: (i) a strain field to remove the valley degeneracy and spatially split the electrons of different valleys to opposite directions transverse to the current flow. By matching the Fermi energy and the gauge potential caused by strain field, we can achieve a situation of no overlap in spatial (angular) space between the transmission angles of the two valleys; (ii) a magnetic field to select the desired valley by transverse shifting of the propagation direction of the transmitted electrons so that electrons of the undesired valley will be totally reflected. We integrate these two effects into a single barrier region which is perfectly transparent for the selected valley, and totally opaque for the other. The proposed system is illustrated in Fig. 1; the uniform uniaxial strain field is applied in the barrier region, while a magnetic barrier is generated by means of two $\delta$-function magnetic field created by ferromagnetic stripes situated at the boundaries of the barrier region.

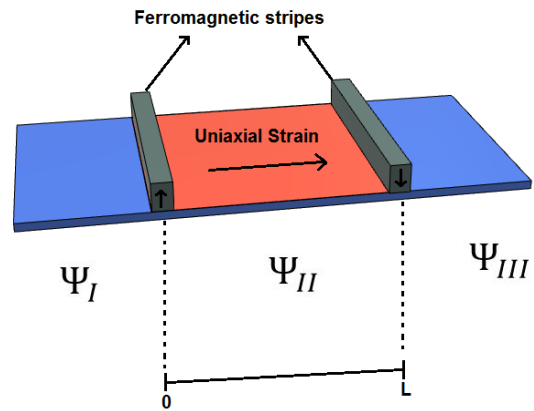

**FIG. 1.** The illustration of the proposed graphene-based valley filter system. The blue color region represents the unstrained monolayer graphene sheet. The red color region is the barrier region under uniaxial strain applied along $\hat{x}$-direction. The electron wave functions for three regions are represented by $\Psi_I, \Psi_{II}, \Psi_{III}$. The arrows on the ferromagnetic stripes show the direction of their magnetization.

The low energy Hamiltonian of graphene under uniaxial strain in the $x$-direction and the magnetic field can be written by

$$H = v_F \hbar \, \vec{\sigma} \cdot (\vec{k} + \eta (v_F \hbar)^{-1} \vec{A}_S + \alpha (v_F \hbar)^{-1} \vec{A}_B) \qquad (1)$$

where, $\eta = \pm 1$ corresponding to the valleys $K$ and $K'$, $\vec{\sigma} = (\sigma_x, \sigma_y)$ are the Pauli matrixes, $\alpha = \pm 1$ representing the direction of the applied magnetic field. Since the strain field affects the local hopping energy $t \to t + \eta \, \delta t_S$ ($t \approx$ 3 eV)[6], it can be described by a gauge potential $\vec{A}_S = \delta t_S \, \hat{y}$. Another gauge potential is due to the delta magnetic field $\vec{B}_z$ created by the ferromagnetic stripes whose magnetization are anti-symmetric, and it is given by $\vec{A}_B = B_0 \, l_B [\Theta(x) - \Theta(x - L)] \hat{y} \equiv \delta t_B$. The incident, propagating and transmitted wave functions shown in Fig. 1 can then be written as

$$\Psi_I(x) = e^{ik_x x} \begin{pmatrix} 1 \\ e^{i\phi} \end{pmatrix} + r \, e^{-ik_x x} \begin{pmatrix} 1 \\ e^{-i\phi} \end{pmatrix},$$

$$\Psi_{II}(x) = a e^{iq_x x} \begin{pmatrix} 1 \\ e^{i\theta} \end{pmatrix} + b \, e^{-iq_x x} \begin{pmatrix} 1 \\ e^{-i\theta} \end{pmatrix},$$

$$\Psi_{III}(x) = t e^{ik_x x} \begin{pmatrix} 1 \\ e^{i\phi} \end{pmatrix}.$$

In the above, the wave vectors can be found by solving the Hamiltonian in Eq. 1. The magnetic field and uniaxial strain along $\hat{x}$-direction both induces a gauge potential in the $\hat{y}$-direction. Therefore, the $\hat{y}$-component of the Fermi wave vector within the barrier, is given by

$$k_y = \frac{\sin\theta \, E + \eta \delta t_S + \alpha \delta t_B}{v_F \hbar}. \qquad (2)$$

where $\theta$ is the the propagation angle of electrons within the barrier. The right-hand side of this equation must be equal to the incident $k_y$ outside of the barrier, due to the conservation of the transverse momentum. By using the relation, $E \sin\phi = E \sin\theta + \eta \delta t_S + \alpha \delta t_B$, where $\phi$ is the incident angle with respect to the $x$-axis, and $E$ is the electron energy, the angle of propagation of the transmitted electrons can be obtained as a function of $\phi$:

$$\theta = \sin^{-1}\left(\frac{E \sin\phi + \eta \delta t_S + \alpha \delta t_B}{E}\right). \qquad (3)$$

Thus, it is clear that the strain and magnetic fields directly affect the transmission angle of the electrons. Finally, the valley-dependent transmission probability $T^\eta(\phi)$ of the system can be calculated by considering the wave function continuity at the two interfaces of the barrier.

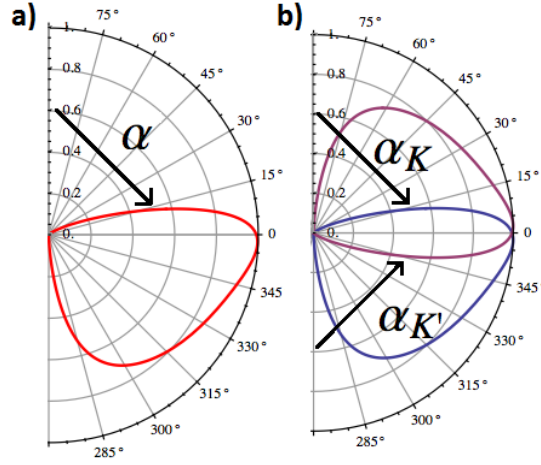

**FIG 2.** The angular dependence of transmission probability $T^{\eta}(\phi)$ in the case of the Fermi energy $E_F = 25$ meV, and barrier width $L = 100$ nm. (a) Shows the effect of magnetic field $\delta t_B = 15$ meV on the transmission probability for valley degenerate system. The angular shift due to the magnetic field is shown by $\alpha$. (b) shows the angular shift caused by the strain applied in the $\hat{x}$-direction. The different valleys experience opposite deflection since strain breaks the valley degeneracy, and has an opposite effect on the two valleys.

**III. Transmission and polarization of the system**

To provide an understanding of the propagation of electrons within the barrier, we analyze the two effects of magnetic field and strain on the transmission separately. If one considers an incident electron with angle $\phi$, it is straightforward to calculate the shift on the angle space by using Eq. 3, which gives the propagation angle of electron in the barrier region. Eq. 3 can be also used to calculate the angle shift due to the barrier since the equation is identical to the Snell's law in optic, which is also valid for electron waves propagating across an interface. Consider an electron which has incident angle $\phi = \pm \pi/2$, and thus, from Eq. 3, its angle after transmission is given by $\sin^{-1}\left(\frac{\pm E + \eta \delta t_S + \alpha \delta t_B}{E}\right)$. The angular shift can be attributed to the transverse Lorentz displacement induced by the magnetic field, and its numerical value can be calculated by substituting the value for $\alpha \delta_B$. Fig. 2 (a) shows this angle shift numerically, which agrees with the above analytic prediction. Similarly, angular shift due to the strain can also be calculated by Eq. 3, but this time direction of the shift depends on the valley spin $\eta$. From Eq. 3, the propagation angle of electrons belonging to different valleys would experience opposite deflection [see Fig. 2 (b)]. In this system, we note that the strain and magnetic fields have antiparallel (collinear) effects on the propagation direction of electrons for the $K'$ ($K$) valley, as shown in Fig. 2. This means that there is no net gauge potential on the desired valley's ($K'$) electrons, which can travel on without any deflection or reflection. For the other valley ($K$) electrons, the sum of the strain and magnetic gauge potentials causes an almost total reflection. Thus, the system does not yield a neutral valley current, and reaches maximum (almost zero) conductivity for the $K'$ ($K$) valley. Note that as in previous treatments [9, 17], we have neglected the effect of the small Zeeman splitting.

To investigate the valley polarization for the whole system shown in Fig. 1, the conductance is first calculated by

$$G_{K(K')} = G_0 \int_{-\pi/2}^{\pi/2} \cos\phi \ T^{K(K')} d\phi,$$

where

$$G_0 = \frac{2e^2}{h}(E_F L_y / v_F \hbar).$$

The polarization value can then be obtained via the conductance values. For instance, the $K$ valley polarization is given by

$$P_K = \frac{G_K - G_{K'}}{G_K + G_{K'}}.$$

Now, let us analyze the polarization of the system by considering varying magnetic and strain field for the case of barrier width $L = 100$ nm and Fermi energy $E_F = 25$ meV. Fig. 3 shows the polarization of the $K$ valley, which approaches unity with increasing $\delta t_S$ and $\delta t_B$. Interestingly, highest polarization is obtained when the gauge potential of magnetic field and strain, also the Fermi energy are equal to each other. Moreover, this perfect polarization is maintained even one or both of the magnetic and strain potentials exceed the Fermi energy. This is not the case, however, for the conductance behavior. As shown in Fig. 3 (b), the conductance reaches a maximum only under the situation of equal magnetic and strain effect. This effect can be understood analytically. Consider a magnetic and strain field of equal strength, i.e., $|\delta t_S| = |\delta t_B|$. For a given magnetic field direction, the gauge potential of magnetic field would have the same sign and magnitude with one of the valleys, but of opposite sign with the other. For the latter, the two gauge fields cancel one another, and it can be readily seen from Eq. 3. that $\theta = \phi$, which means perfect transmission for the selected valley (without any deflection). On the other hand, the same configuration yields an argument of arcsin that has a magnitude greater than 1, which corresponds to a transmission angle outside of the allowable range of $-\frac{\pi}{2} < \theta < \frac{\pi}{2}$ for the other valley (for which the strain and magnetic potentials have same sign). Hence, this valley spin will be totally reflected giving rise to maximum valley polarization at the same time. Note that if there exists an inequality between $\delta t_S$ and $\delta t_B$, this translates to an inequality between the incident $\phi$ and transmitted $\theta$, thus resulting in a decrease in the conductance.

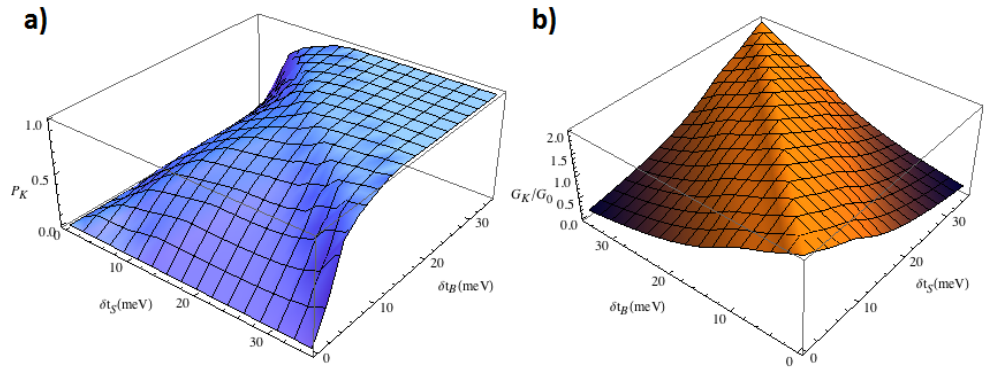

**FIG. 3.** (a) The polarization of the K valley in the case of $E_F = 25$ meV, $L = 200$ nm, when the gauge potentials $\delta t_B$ and $\delta t_S$ caused by strain and magnetic field varied between 0 and 30 meV. (b) The conductance of the system for the same configuration.

Numerically, we obtained a valley polarization exceeding 99% for the configuration $|\delta t_S| = |\delta t_B|$, and this value can be attained for both valleys. Compared to the previous work in Ref. 9, where a similarly high value of valley

polarization is achieved, there is a marked difference in the transmitted conductance. In the system proposed in Ref. 9, the conductance under the situation of 99% valley polarization is only $G_K = 0.31 G_0$, which is very low compared to that achieved in the present system, which is $G_K \approx 2 G_0$ (i.e., almost perfect conductance). Moreover, unlike the previous works, the valley filtering efficiency of our system can be made to occur at any specific Fermi energy. To obtain the almost perfect valley filter operation, it is sufficient just to have the magnetic and uniaxial strain potentials to match with the Fermi energy. This makes the system more versatile and practical. It also the experimental feasibility of our proposed design, since one can potentially attain simultaneously high valley polarization and conductance with low magnetic and strain potential. In our numerical examples, we have assumed parameter values of $E_F = 25$ meV, $\delta t_B = \delta t_S \leq 25$ meV, which can be realized with current technology by applying magnetic field $B_0 \sim 1$ T and ($\approx 0.8\%$) strain. These values have already been demonstrated experimentally[19-21].

**IV. Conclusion**

We have shown analytically and numerically the generation of pure valley-polarized current by means of a single strained barrier region in monolayer graphene, whose boundaries are demarcated by localized fringe magnetic fields. Unlike previously proposed valley filters, high valley polarization is not confined to only particular values of strain and magnetic fields, and does not come at the expense of high conductance. In our proposed system, the only requirement for simultaneously achieving high valley filtering efficiency and high conductance is by matching both the strain and magnetic field gauge potentials to the Fermi energy. Besides, the use of a single strain barrier system makes the application easy to realize experimentally. We also discuss the underlying principle of the valley filtering, which is based on (i) using the uniaxial strain to lift the valley degeneracy and shift the perfect transmission angles of the incident electrons to opposite transverse directions for different valleys, and (ii) applying a matching magnetic field to shift the whole transmission profile due to the transverse Lorentz displacement. This displacement serves to block one of the valley spins from the transmission range (by total reflection), while aligning the other valley to the initial position (thus guaranteeing maximum transmission). Thus, the barrier becomes transparent for the electrons coming from the desired valley. As a consequence, we have showed that pure valley polarization coupled with high conductance can be achieved with experimentally feasible parameters.

**Acknowledgments**

The authors would like to thank the MOE Tier II grant MOE2013-T2-2-125 (NUS Grant No. R-263-000-B10-112), and the National Research Foundation of Singapore under the CRP Program "Next Generation Spin Torque Memories: From Fundamental Physics to Applications" NRF-CRP9-2013-01 for financial support.